\documentclass[runningheads]{llncs}
\usepackage[colorlinks=true,pagebackref]{hyperref}

\usepackage{graphicx}
\usepackage{mathtools,xcolor}
\usepackage{xcolor}
\usepackage{multirow,bbm,float}
\usepackage{amssymb}
\usepackage{pifont}
\usepackage{algpseudocode}
\usepackage{array}
\newcommand{\cmark}{\ding{51}}%
\newcommand{\xmark}{\ding{55}}%

\begin{document}
\title{KS-GNNExplainer: Global Model Interpretation Through Instance Explanations On Histopathology images}
\titlerunning{KS-GNNExplainer}
%
\author{Sina Abdous, Reza Abdollahzadeh, Mohammad Hossein Rohban}
\authorrunning{Abdous et al.}


%

\institute{Sharif University of Technology \\
\email{\{sina.abdous,re.abd,rohban\}@sharif.edu}}

\maketitle              
\begin{abstract}
Instance-level graph neural network explainers have proven beneficial for explaining such networks on histopathology images. However, there has been few methods that provide model explanations, which are common patterns among samples within the same class. We envision that graph-based histopathological image analysis can benefit significantly from such explanations. On the other hand, current model-level explainers are based on graph generation methods that are not applicable in this domain because of no corresponding image for their generated graphs in real world. Therefore, such explanations are communicable to the experts. To follow this vision, we developed KS-GNNExplainer, the first instance-level graph neural network explainer that leverages current instance-level approaches in an effective manner to provide more informative and reliable explainable outputs, which are crucial for applied AI in the health domain. Our experiments on various datasets, and based on both quantitative and qualitative measures, demonstrate that the proposed explainer is capable of being a global pattern extractor, which is a fundamental limitation of current instance-level approaches in this domain.

\end{abstract}
\section{Introduction}

Recent advances in machine learning, and particularly deep learning, have transformed histopathological image understanding. However,
such advantages come at the cost of a reduced transparency
in decision-making processes \cite{28}. Given the importance of reasoning in any clinical decision, pathologists often seek explanations for the deep learning decisions. Therefore, inspired by the explainability techniques on natural images \cite{57}, several explainers have been implemented
in the digital pathology, such as feature attribution based methods \cite{10} but, inter-entity interactions are ignored using these methods,  which is destructive in the pathologist's diagnosis process.


To address the aforementioned concerns, a histological image is represented as an entity graph, with nodes and edges indicating biological entities and inter-entity interactions, respectively to make explainability possible in the entity space, such as cells. The graph is then used to form a Graph Neural Network (GNN) \cite{35}, which aims to solve the graph classification problem. Subsequently, graph explainers \cite{7} are employed, which could highlight the responsible entities for the final diagnosis, providing pathologists with intuitive explanations.

\medskip

Recently several graph explainers have been proposed, which can be classified into two categories: instance-level  and model-level explanations \cite{DBLP:journals/corr/abs-2012-15445}.
 Instance-level explanations  focus on explaining the model prediction for a given graph instance, which means their inability to consider a batch of samples and extract common patterns like tumor subregions among similar samples. Meanwhile, current model-level explanation methods aim at understanding the general behavior of the model using graph generation methods, which is not applicable in the context of explaining GNNs for histopathology images, as there is no corresponding image for the generated graphs in real world.

In this paper, first, a framework to compare explainers for the histopathology images is presented. Then, based on our findings from the proposed framework, we present a novel model-level explainer, which addresses the limitations of methods in both categories. The core idea of the proposed benchmark framework is to gradually remove the nodes that are declared as the most important ones in making the decision by an explainer, and assess whether the model output changes in response to such removals. By doing this, we get a cumulative distribution on the number of instances that their predicted labels change. Repeating the same procedure for the least important nodes, we get a second cumulative distribution. We then propose to use the Kolmogorov-Smirnov (KS) test on the two empirical cumulative distributions to quantify the effectiveness of the explainer. One would expect a shift between these two distributions in case that the most and least important nodes have been correctly identified. We show that our validation framework is sound in the sense that the explainers with high proposed KS score could be leveraged to enhance the diagnosis accuracy much better than other explainers.

Then, we present KS-GNNExplainer which is built upon GNNExplainer, a competent instance-level explainer of GNNs \cite{DBLP:journals/corr/abs-2011-12646}, and extend it to be a model-based explainer by incorporating (1) pairwise embedding similarities among explanation graphs with the same label; and (2) the mentioned KS-score in its objective function. Here, we seek subgraphs, one in each instance of a batch of data, with similar structures (global model explainers) that optimize the KS score, once removed from the graphs.
We show the effectiveness of our approach on a variety of datasets.


In summary, our key contributions are as follows:
\begin{enumerate}
    \item{ Introducing a simple method based on two sample Kolmogorov-Smirnov (KS) test for benchmarking graph neural networks explainers on histopathology images.}
    \item{ Introducing a model-level explainer (KS-GNNExplainer), which distinguishes itself from all previous works by introducing a novel objective function that suits histopathology images.} 
\end{enumerate}

\begin{figure}[H]
	\centering
	\includegraphics[scale=0.08]{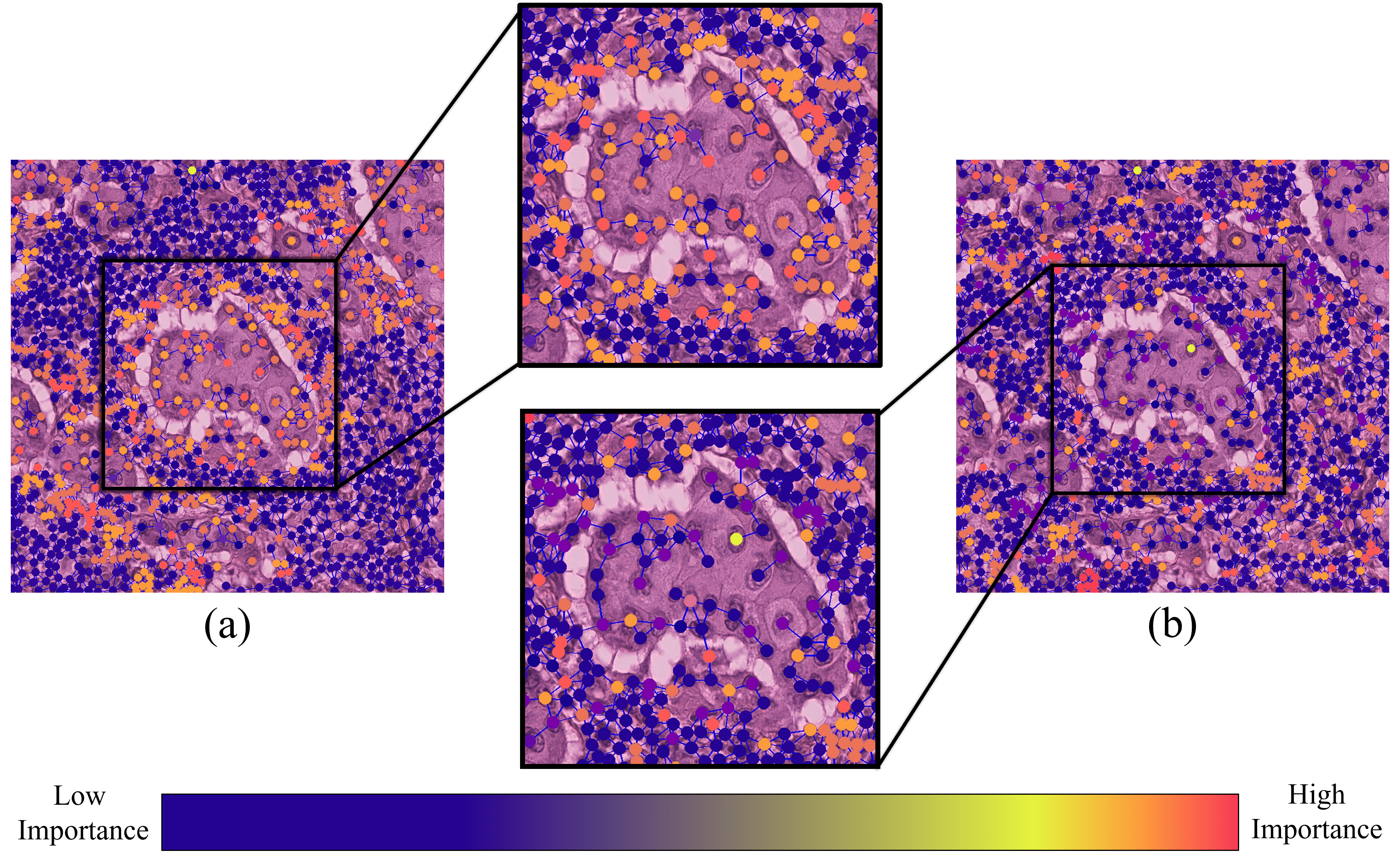}
	\caption{Explanation graph produced for a ductal carcinoma in situ RoI by (a) KS-GNNExplainer and (b) GNNExplainer as current state-of-the-art method \cite{DBLP:journals/corr/abs-2011-12646}. Nucleis involved in lymph-vascular invasion (confirmed by a pathologist) are detected in a more integrated appearance using our proposed method.}
	\label{fig:Qualitative}
\end{figure}


\section{Background}

\subsection{Notations}

Cell-graphs (CG) created from the pathology images are defined as an undirected graph $G_{\mathrm{CG}}=(V, E, H)$ composed of  vertices $V$, and edges $E$. An embedding $h \in \mathbb{R}^{d}$ describes each vertex, which is collectively  expressed in its matrix form as $H \in \mathbb{R}^{|V| \times d}$. A symmetric adjacency matrix  $A \in \mathbb{R}^{|V| \times|V|}$ describes the graph topology, where $A_{u, v}=1$ if an edge exists between the vertices $u$ and $v$.

\medskip

We use the Hover-Net \cite{https://doi.org/10.48550/arxiv.1812.06499} for nuclei segmentation. The cell graphs are then made based on the segmented nuclei as their nodes. Each node is then connected to its $k$-NNs. The node features are image patches of a fixed size around the nuclei. Finally, HACT-Net \cite{44}, is employed to build fixed-size graph embeddings from the CGs, which are fed into a Multi-Layer Perceptron (MLP) to predict the cancer stage labels. 


\subsection{Metrics}

Here, we used $\text{Fidelity}+^{a c c}$ score \cite{DBLP:journals/corr/abs-2012-15445}, and call it  Fidelity onward, for studying faithfulness of different methods by masking the nodes that are detected as the most important ones in the explanation and expect that the predictions of model would change under the mentioned masking.

\begin{equation}
\text { Fidelity }=\frac{1}{N} \sum_{i=1}^N\left(\mathbbm{1}\left(\hat{y}_i=y_i\right)-\mathbbm{1}\left(\hat{y}_i^{1-m_i}=y_i\right)\right),
\label{eq:fidelity}
\end{equation}

\begin{figure}[t]
\centering
 \begin{center}
\begin{tabular}{ c c c }
  \includegraphics[width=0.32\linewidth]{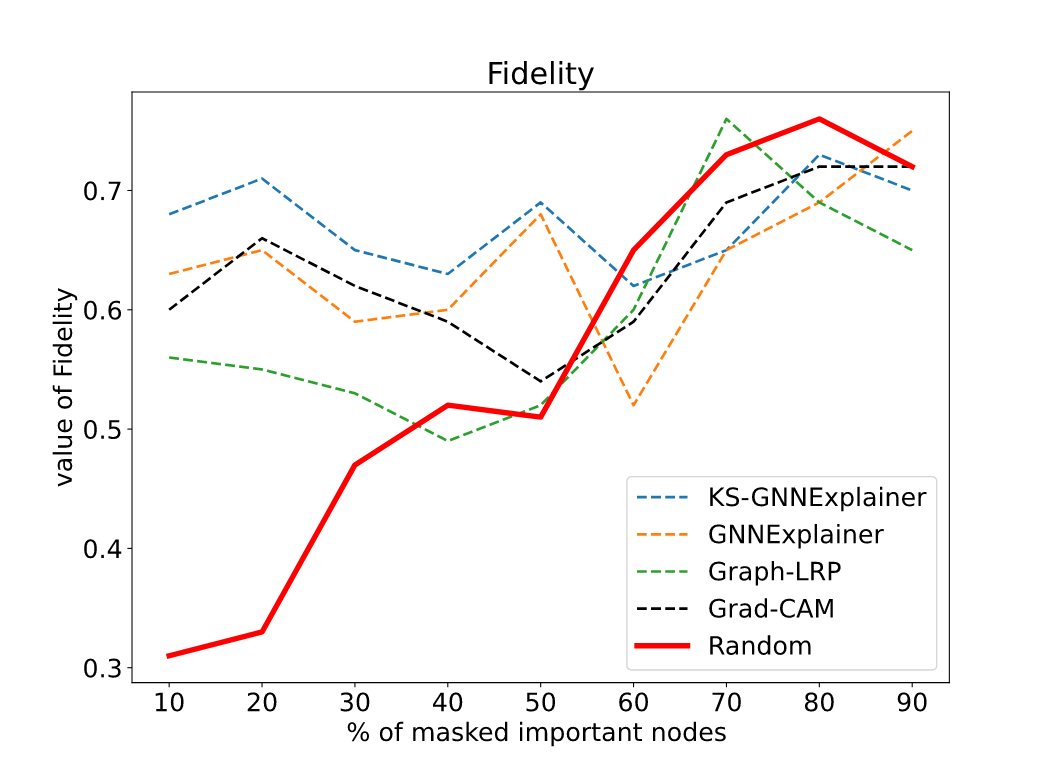} &  \includegraphics[width=0.32\linewidth]{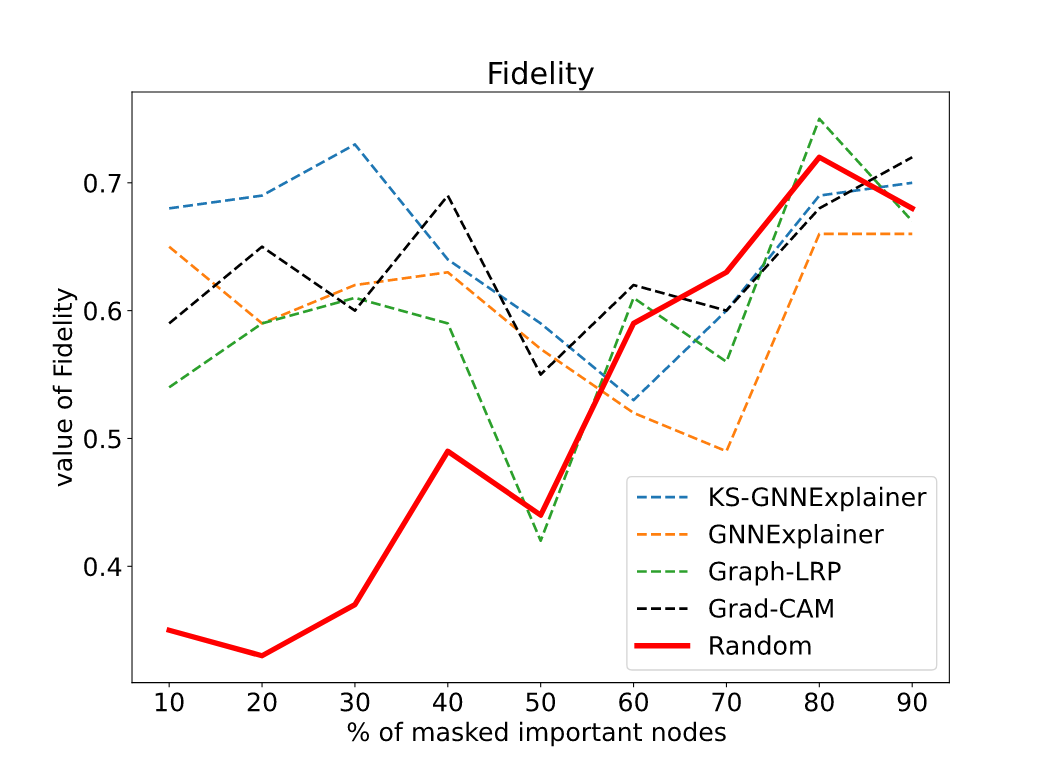} &  \includegraphics[width=0.32\linewidth]{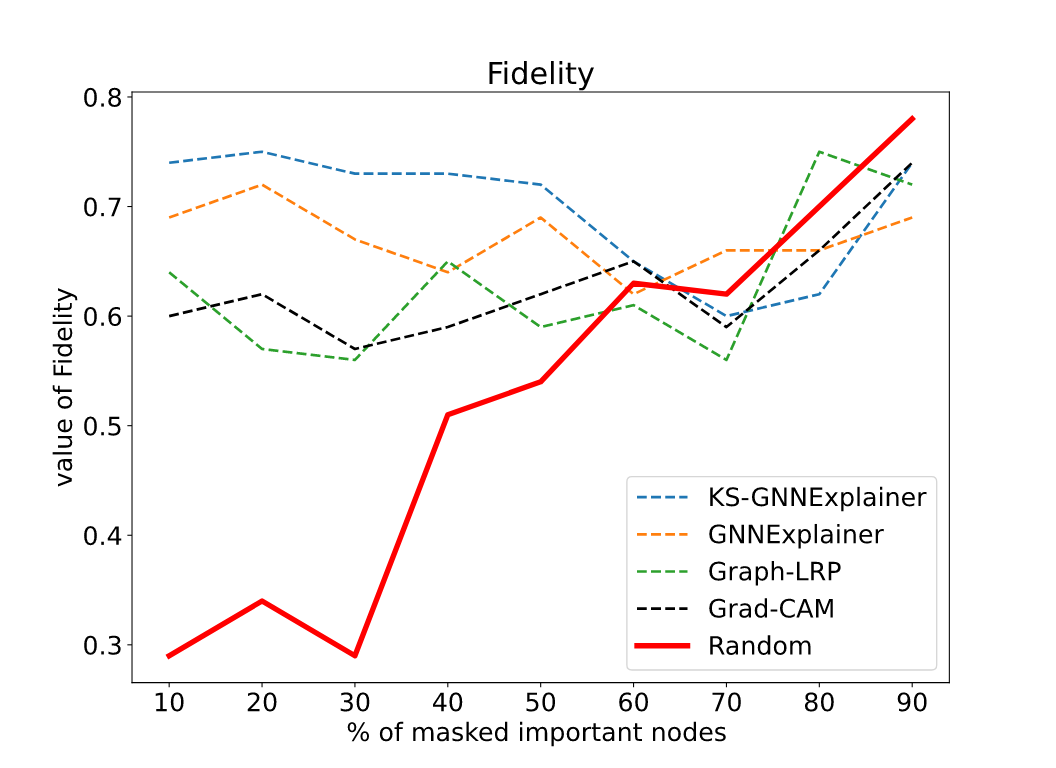}\\ 
 (a) & (b) & (c)    
\end{tabular}
\end{center}
	\caption{Performance of studied explainers in addition to random explainer using Fidelity on (a) BRACS, (b) BACH, and (c) CRC. The effectiveness of KS-GNNExplainer on masking up to nearly 50\% of important nodes can be demonstrate.}
	\label{fig:fidelity}
\end{figure}

Examining Eq. \ref{eq:fidelity}, $y_i$ is original prediction of the $i$-th input graph, and $N$ is the total number of input graphs. Furthermore, $\hat{y}_i^{1-m_i}$ denotes prediction of $i$-th graph after masking its $m_i$ most important nodes. Higher values for Fidelity indicate more discriminative features being identified. As the generated importance map for nuclei distribution using studied explainers are continuous, we plot the Fidelity score based on different thresholds of importance in Fig. \ref{fig:fidelity}.

\section{KS-Bench framework}
As the first step, we propose a new framework to benchmark explainers on the histopathology images. Unlike previous work \cite{pmlr-v156-jaume21a}, our benchmark evaluates explanation qualities, which is then used to improve the classification. For this purpose, we have considered graph-based explainers such as GNNExplainers \cite{10.5555/3454287.3455116}, GraphLRP \cite{DBLP:conf/textgraphs/SchwarzenbergHH19}, Grad-CAM \cite{8237336}, and Grad-CAM++ \cite{8354201}.

The main idea in KS-Bench is the gradual removal of the most and the least important vertices that are reported by each explainer, and then observing changes of the output label of the model. Note that unlike  the image inputs, node and edge removal often do not result in artifacts and distribution shifts in the input graphs, and hence does not prohibit the use of the original GNN model in assessing the prediction of masked inputs.
This is also due to the fact that the high variety of graph structures and sizes are used in the training phase. 
\begin{table*}[ht]
\caption{The KS statistic, p-value, and class accuracies of HACT-Net when incorporating the explainer importance score in the top 30\% nodes in the BRACS dataset. Note that the first row is baseline accuracy of HACT-Net on BRACS dataset.}
\centering
\begin{center}
\resizebox{\columnwidth}{!}{%
\begin{tabular}{ c | c  c | c c c c c c c}
\hline
Explainer  & KS value & p-value & Normal  & Benign  & UDH  & ADH   & FEA  & DCIS  & Invasive   \\
\hline
None    & - & - & 60.2\% & 53.3\% & 52.5\% & 51.4\% & {\bf 65.1}\% & 56.2\% & 67.7\% \\
\hline
GNNExplainer & {\bf 0.748} & $2.04\mathrm{e}{-10}$ & {\bf 68.2} \% & {\bf 58.5}\% & {\bf 56.1}\% & 59.6\% & 63.7\% & {\bf 68.9}\% & {\bf 72.4}\% \\
GraphLRP    & 0.552& {\bf $2.17\mathrm{e}{-12}$} & 57.9\% & 52.0\% & 54.1\% & 54.2\% & 56.1\% & 60.2\% & 64.9\%\\
Grad-CAM    & 0.615& $2.21\mathrm{e}{-7}$ & 63.5\% & 54.1\% & 54.7\% & 57.9\% & 58.2\% & 64.7\% & 65.4\%\\
Grad-CAM++  & 0.658& $1.93\mathrm{e}{-8}$ & 66.9\% & 57.2\% & 55.3\% & {\bf 61.0}\% & 60.8\% & 65.3\% & 68.1\%\\
\hline
\end{tabular}
}
\end{center}
\vskip -0.1in
\label{tab:ks-bench}
\end{table*}
More specifically, we  remove the most, and the least important nodes reported by each explainer in two separate phases, gradually (from 0\% to 100\% by 5\% increments). Then, we obtain two cumulative distributions for each explainer, corresponding to each phase. At any percentage point $j$, these distributions represent the proportion of the inputs that has experienced any change in the model output label after masking lower than $j$ percent of their nodes so far. We envision KS statistics compared to existing metrics like Fidelity, gives a higher discrimination power to evaluate explainers as it checks the effect of masking both most and least important components simultaneously while these two type of components are assessed separately in the case of Fidelity ($\text{Fidelity}+$ and $\text{Fidelity}-$), which makes it hard to interpret if only either of the metrics improves. 
We used the two-sample Kolmogorov–Smirnov (KS) test to compare these two cumulative distributions, which will assign a KS value to each explainer. 
Using KS test, the ideal explainer is the one that the prediction of the trained
model changes for a significant number of sample images after removing first few most important nodes. Also, the label is expected to remain almost unchanged for most of images by removing the least important nodes. 

\medskip

To validate our benchmarking scheme,
we incorporate the node importance score given by each explainer as a feature in the HACT-Net to potentially improve the accurate classification. We then correlate the performance of the modified HACT-Net with the proposed KS score. More specifically, in the modified HACT-Net, we extend the node features to include a flag of whether that node is among the 30\% of most important nodes. 
The results are reported in Table \ref{tab:ks-bench}. 
An important observation here is that GNNExplainer achieves the highest KS value, and also its importance scores improve the HACT-Net by a significant amount in most classes. We see a similar trend in the Grad-CAM++, which is the runner up according to the KS values. This observation shows that our KS benchmark gives an indication on the usefulness of an explainer in improving the classification .




\section{KS-GNNExplainer}

\begin{figure}[H]
	\centering
	\includegraphics[scale=0.3]{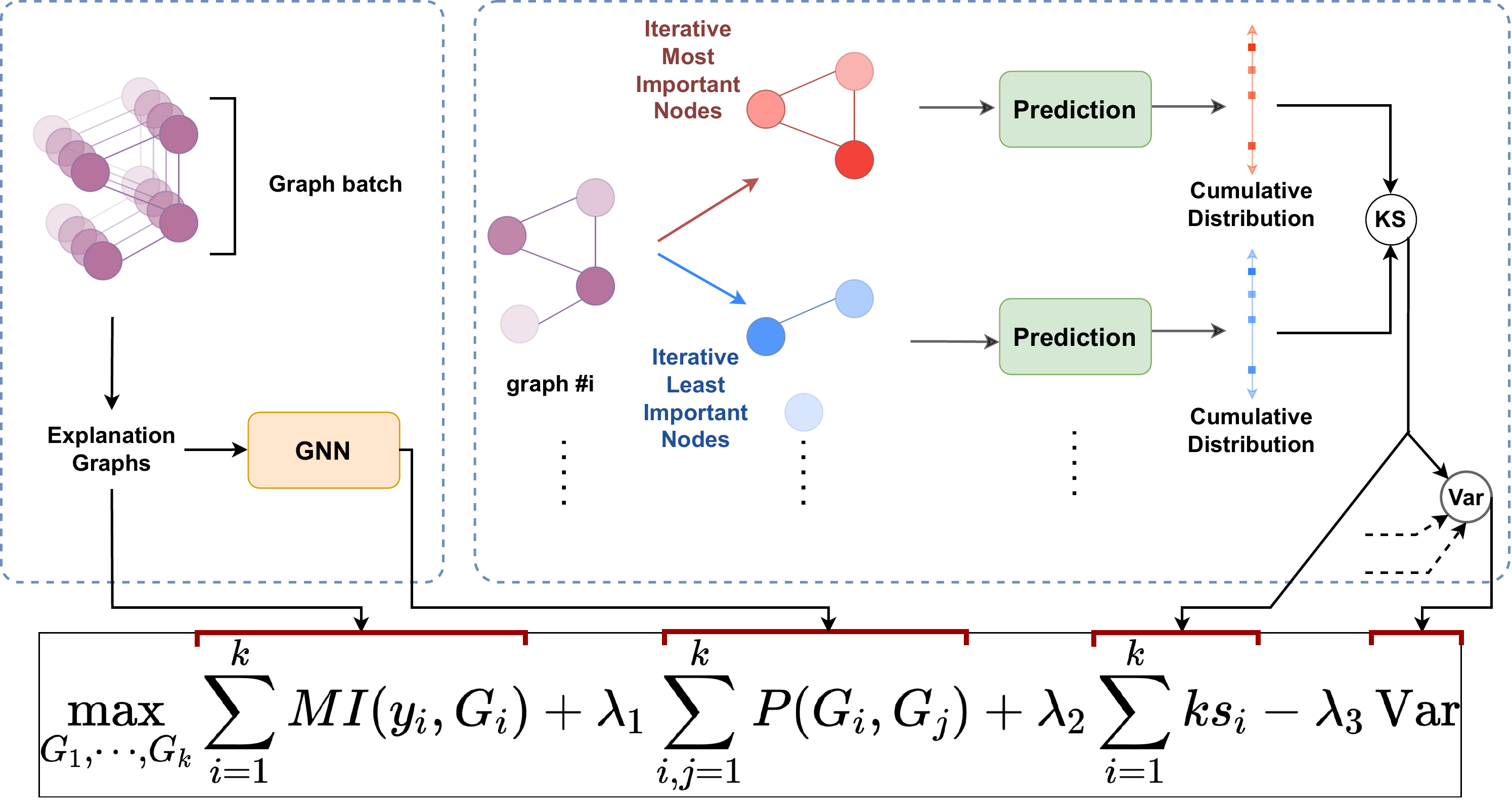}
	\caption{The objective function of KS-GNNExplainer. It  maximizes three components (mutual information, pairwise similarity, and summation of KS value) among all samples, in addition to minimizing the variance of obtained KS values.}
	\label{fig:objective-func}
\end{figure}

Based on our findings from proposed benchmark in previous section, we reformulate the objective function of GNNExplainer as our baseline method to interpret
the whole model while maintaining its simplicity as an instance-based explainer. GNNExplainer as a perturbation-based method, tries to find an explanation subgraph which maximizes mutual information between the true label and the subgraph using and iterative mask generation algorithm.So for the task of generating
explanation for a single graph, we considered a batch of graphs with the same label, and leveraged the
objective function from a simple optimization problem which maximizes mutual information between the explanation subgraphs and label to a general optimization problem as illustrated in Fig. \ref{fig:objective-func}.
\vspace{-0pt}
\begin{gather}
\max _{G_1, \cdots, G_k} \sum_{i=1}^k MI\left(y_i, G_i\right) + 
\lambda_1 \sum_{i,j=1}^k P\left(G_i, G_j\right) +
\lambda_2\sum_{i=1}^k {ks}_i -
\lambda_3 Var,
 \label{eq:objective-func}
\end{gather}
where $Var=variance({ks_1,\cdots, ks_k})$ and KS value for every graph in each iteration is calculated using two-sample KS test 
and is as follows:
\begin{equation}
ks_i=\sup _x\left|F_{1, n}(x)-F_{2, m}(x)\right|,
 \label{eq:ks}
\end{equation}
where $n_1$ and $n_2$ are observations, and $F_{1,n}$ and $F_{2,m}$ are the empirical cumulative distribution functions of the first and the second distributions, respectively.
Examining Eq. \eqref{eq:objective-func}, it maximizes summation of four terms. First term is sum of mutual information for all samples in the batch, which is the main idea of GNNExplainer at the single instance level. Continuous relaxation suggested in \cite{10.5555/3454287.3455116} has applied here to overcome the discrete optimization difficulty. Inspired by objective function of \cite{100}, the second term is sum of pairwise similarity for each two subgraphs, based on their cosine similarity of their embeddings from the HACT-Net model, denoted by $P(., .)$. This term ensures that explanation subgraphs of inputs from the same class are similar, which is step towards a model-level explanation. The third term is sum of KS value for each subgraph in each step. Our objective function also tries to minimize variance between all KS values as we would expect a uniformly large KS values across samples of batch. 

$\lambda_1$, $\lambda_2$, and $\lambda_3$ are coefficients that determine the relative importance of different terms, and optimal values for them are empirically set using the validation set of the BRACS. 

\section{Experiments}

\subsection{Datasets}
We focus on prediction of the Breast cancer subtypes. So we  evaluate our framework on public datasets, BRACS \cite{https://doi.org/10.48550/arxiv.2111.04740} and BACH \cite{ARESTA2019122}, as well as a nuclei annotated dataset, BreCaHAD \cite{aksac2019brecahad}. To further examine our method, we used Colorectal Cancer Grading (CRC) dataset \cite{Awan2017} as well.

\begin{figure}[H]
\centering
\includegraphics[scale=0.14]{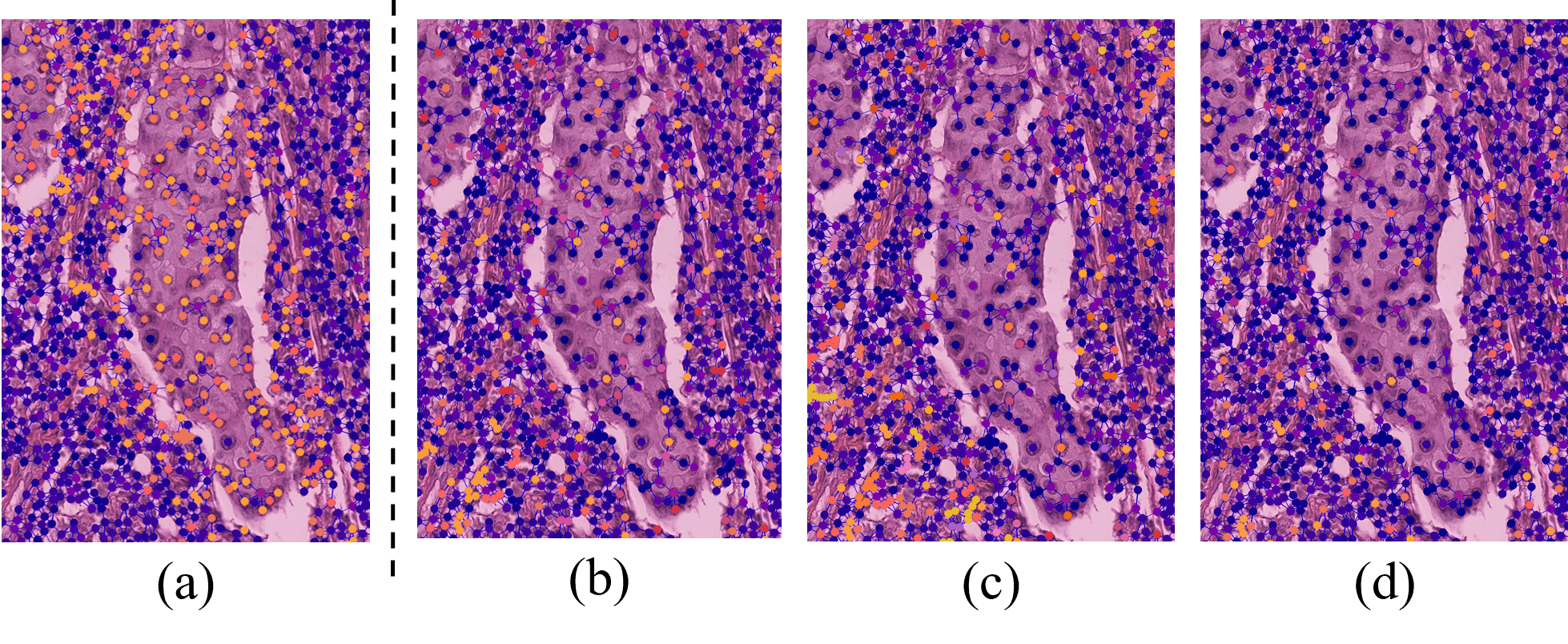}
\caption{Qualitative results on ductal carcinoma in situ RoI sample of BRACS. Darker colors correspond to lower importance. Each column correspond to an explainer: (a) KS-GNNExplainer, (b) GNNExplainer, (c) Grad-CAM, and (d) GraphLRP.}
\label{fig:Qualitative1}
\end{figure}

\subsection{Qualitative analysis}
We conduct a set of experiments to determine how each explainer performs in recognizing lymph-vascular invasion as a critical sign in cancer grading as well as discriminating tumor from non-tumor nucleis. Fig. \ref{fig:Qualitative1} shows the nuclei importance map of studied explainers.
It shows effectiveness of our method over others in detecting lymph-vascular invasion as a critical sign of DCI subtype of breast cancer. We envision the superior performance in extracting common patterns among instances is due to our model-level objective function through applying both pairwise similarity and KS test.

\subsection{Quantitative results}
We also use the mentioned metrics to further analyze previous explainers quantitatively. First, we plot changes in the Fidelity for all explainers, as well as a random one, for different thresholds on each dataset in Fig. \ref{fig:fidelity}. As can be seen, superior performance of the KS-GNNExplainer over other methods is evident.

In addition, the macro-averaged F1 score results are shown in Table \ref{tab:macro-f1}. Here, we considered a binary classification of tumor vs. non-tumor nuclei through the importance score. We assign nuclei with the importance score of 0.5 and greater as tumor, and else as non-tumor, and compare against the ground truth nuclei label. Here, our KS-GNNExplainer method shows superior macro averaged F1 score compared to the other explainers.

\begin{table}[t]
\caption{The macro F1 score for the nuclei classification on BreCaHAD.}
\label{tab:KS statistic}
\vskip 0.15in
\begin{center}
\begin{small}
\begin{sc}
\begin{tabular}{l|cc|ccr}
\hline
Explainer & \multicolumn{2}{c|} {per-class F1 score} & Macro-averaged f1 score \\
\multicolumn{1}{c|}{}  & Tumor & non-Tumor     &       &      \\
\hline
KS-GNNExplainer    & {\bf 0.725}  & 0.669 & {\bf0.697} \\
GNNExplainer    & { 0.692} & {\bf 0.682} &  0.687\\
GraphLRP        & 0.426& 0.553 & 0.489\\
Grad-CAM         & 0.595& 0.503 &  0.549\\
\hline
\end{tabular}
\end{sc}
\end{small}
\end{center}
\vskip -0.1in
\label{tab:macro-f1}
\end{table}

\subsubsection{Ablation study:}
We conduct a thorough study to show how each term contributes to our objective function. We draw the following observations from Table \ref{tab:ablation}: (1) Considering our objective function, by removing the KS test part (third and fourth terms), a marginal decrease in Fidelity score can be seen and is a proof of our effective method (Exceptions like the one for BRACS, are a fine case study for future works); (2) The Fidelity score dropped slightly by removing the second term, shows the idea of making the embeddings of subgraphs belonging to the same subclass is also beneficial. Our findings indicate that unifying all terms in our method, significantly leverages the GNNExplainer by through extracting common patterns from samples.

\begin{table*}[ht]
\caption{Ablation study on different components of the proposed objective function.}
\vskip 0.15in

\centering
\begin{center}
\begin{tabular}{c |c c c c c c }
\hline
Dataset  & $term_{MI}$  & $term_{P}$  & $term_{KS\ sumation}$    & $term_{KS\ variance}$& Fidelity score  \\ 
\hline
\multirow{4}{*}{BACH}     & \cmark & \cmark & \cmark  & \cmark  & 0.68 \\
\cline{2-6}
 & \cmark & \xmark & \xmark & \xmark  &  0.32 ($\downarrow$  0.36 ) \\
 & \cmark & \cmark & \xmark & \xmark  &  0.49 ($\downarrow$  0.19) \\
    & \cmark  & \cmark &  \cmark & \xmark  &  0.52 ($\downarrow$ 0.16)\\
    & \cmark & \xmark &  \cmark & \cmark  &  0.56 ($\downarrow$ 0.12)\\
\hline
\multirow{ 5}{*}{BRACS}     & \cmark & \cmark & \cmark  & \cmark  & 0.63  \\
\cline{2-6}
 & \cmark & \xmark & \xmark & \xmark  &  0.28 ($\downarrow$  0.35 ) \\
 & \cmark & \cmark & \xmark & \xmark  &  0.40 ($\downarrow$  0.23) \\
    & \cmark  & \cmark &  \cmark & \xmark  &  0.67 ($\uparrow$ 0.04)\\
    & \cmark & \xmark &  \cmark & \cmark  &  0.45 ($\downarrow$ 0.18)\\
\hline
\end{tabular}
\end{center}
\vskip -0.1in
\label{tab:ablation}
\end{table*}

\section{Conclusion}
We proposed KS-GNNExplainer, the first model-level graph neural network explainer inspired by instance-level methods. The key contribution of the KS-GNNExplainer arises from the insight that current instance-level explainers, lack a comprehensive approach to look for similar information in different samples simultaneously without examining the model directly. We have proved its efficiency through our experiments on different datasets.

\bibliographystyle{splncs04}
\clearpage\bibliography{reference}

\end{document}